\documentclass[prl,manuscript,aps]{revtex4}

\usepackage{amssymb,amsmath}
\usepackage{graphicx,float}

\begin{document}

\title{ \Large \bf %\flushleft 

On the derivation of power-law distributions within classical  statistical mechanics 
far from the thermodynamic limit}
\author{\large  \flushleft 
Rudolf Hanel$^{1,2}$  and  Stefan Thurner$^{1,}$\footnote{
 {\bf Correspondence to:}\\
 Stefan Thurner  \\Complex Systems Research Group, Medical University Vienna\\
 W\"ahringer G\"urtel 18-20; A-1090 Vienna, Austria\\
 T +43 1 40400 2099 F +43 1 40400 3332; thurner@univie.ac.at
} 
}
\affiliation{ %\flushleft
$^1$ Complex Systems Research Group; HNO; Medical University Vienna; Austria   \\ 
$^{2}$ Institute of Physics; University of Antwerp; Groenenborgerlaan 171; Belgium\\
 \hspace{1cm}
} 
%\date{}
\begin{abstract}
\noindent
We show that within classical statistical mechanics without taking the 
thermodynamic limit, the most general Boltzmann factor for the canonical 
ensemble is a  $q$-exponential function.
The only assumption here is that microcanonical 
distributions have to be separable from of the total system 
energy, which is the prerequisite for any sensible measurement.  
We derive that all separable distributions are parametrized 
by a mathematical separation constant $Q$ which can be  related to 
the non-extensivity $q$-parameter in Tsallis distributions. 
We further demonstrate that nature fixes the separation constant $Q$ to 1 
for large dimensionality of Gibbs $\Gamma$-phase space. 
Our results will be relevant for
systems with a low-dimensional $\Gamma$-space, 
for example nanosystems, comprised of a small number of particles 
or for systems with a dimensionally 
collapsed phase space, which might be the case for 
a large class of complex systems. 
\\

%\vspace{1cm}

%51.30.+i % Thermodynamic properties, equations of state 
\end{abstract}
\maketitle

\section{Introduction}

There is a certain need in the physical, chemical, biological, 
social and economical  sciences to understand the origin and 
ubiquity of power law-distributions. Many of these distributions
appear to be $q$-exponentials upon closer inspection, which are defined as 
\begin{equation}
e_q^x\equiv  [1+(1-q)x]^{1/(1-q)}
\quad ,
\end{equation}
for $1+(1-q)x \ge 0$. 
An appealing attempt for a general approach to the matter is 
to construct a thermostatistics, where with the use of classical 
principles, such as the variational principle and the 
second fundamental theorem of thermodynamics, one 
would be able to naturally derive these distributions. 
The canonical entropy associated to $q$-exponential distribution 
functions is the so-called Tsallis entropy \cite{tsallis88,gellmann}, 
\begin{equation}
S_q\equiv \frac{1-\int d\Gamma \rho^q }{q-1}   \quad , 
\end{equation}
where $\rho$ is the the normalized energy density and $d\Gamma$ 
indicates phase space integration. 
Tsallis entropy 
which is a generalization of Boltzmann-Gibbs (BG)
entropy is in principle  non-additive and non-extensive.  
Classical BG entropy is  recovered in the limit $q\to 1$. 

In contrast to the vast amount of papers in non-extensive statistical 
physics, there has been relatively little effort to derive 
power-law distributions, in particular 
$q$-exponential distributions, from first statistical mechanics
 principles. To mention 
some fruitful work in this direction, 
in \cite{carati} it was shown that if phase space volumes 
are not covered in a Poissonian manner, the resulting 
entropy functional can be of Tsallis type. 
Within the framework of superstatistics \cite{beck}
the inverse temperatures $\beta$  are 
considered to fluctuate, such that a generalized Boltzmann factor,  
$B=\int d \beta f(\beta) e^{-\beta E}$ arises. If $f$ is a $\chi^2$ distribution, 
the natural distribution functions are $q$-exponentials. 
An other approach was taken in \cite{abe}, where the non-uniqueness 
of counting rules are discussed. It is shown that 
a propper modification of the log-counting rule yields Tsallis entropy instead 
of the usual BG entropy.
In a recent paper \cite{us}, we could show that a more general view is 
possible, in the sense  that the most general Boltzmann factor that 
can be derived for the canonical ensemble is exactly a  $q$-exponential. 
This result might be of relevance for systems with a low dimensionality of 
$\Gamma$ phase space (Gibbs). Examples for such systems are 
nanosystems, which have gained recent interest due to being at the edge of 
technical accessibility, and for systems, whose phase space has 
effectively collapsed in dimensions (like a fractal), 
due to long range interactions, 
aging, synchronization, as is the case in many complex systems.

\section{Derivation of the general Boltzmann factor}

The following argument is solely based on the observation that  
any thermodynamic system which can be 
measured in equilibrium must be {\it energy  separable}, i.e.   
thermodynamic  quantities of the measured system must  
not explicitly depend on the energy of the total system, $E$.   
We consider a sample (observed system)
in contact with a reservoir (bath). The energy of the 
sample is $E_1$, the  energy of the reservoir is 
$E_2$, such that the total system has a 
constant total energy
$E=E_1+E_2$. The number of microstates   
are $\omega_1(E_1)$ and $\omega_2(E_2)$ for the sample and 
the bath, respectively.
The energy of the sample fluctuates around its equilibrium (extremal)
value denoted by $E_*$.
Thermal contact of the two systems means that the 
Hamiltonian of the total system is $H=H_{1}+H_{2}$ 
and the partition function $Z(E)$ is the convolution 
of the two microcanonical densities 
\begin{equation}
Z(E) =  \int\limits^{E}_{0} dE_{1} 
                  \omega_{1}(E_{1}) \omega_{2}(E-E_{1}) \quad  {\rm with} \quad 
\omega_{i}(E_{i})=\int d\Gamma_{i}\quad\delta(H_{i}-E_{i}) \quad .
\end{equation}
Following usual reasoning, we pass from the microcanonical
to the canonical description. Its density $\rho$ is given  
(up to a constant multiplicative factor) by
\begin{equation}
\rho(E_{1})=\omega_{1}(E_{1})\omega_{2}(E-E_{1})Z^{-1}(E) \quad .
\label{rho}
\end{equation}
Note, that this description is entirely dictated by the equations of motion. 
Assuming the existence of a unique extremal configuration  
at some $E_{1}=E_*$ defined by $\delta \rho=0$, leads to the definition of 
inverse temperature
\begin{equation}
        \frac{\omega_{1}'}{\omega_{1}}\big|_{E_1=E_*}  =
	\frac{\omega_{2}'}{\omega_{2}}\big|_{E_2=E-E_*}  \equiv \beta
        \frac{1}{k_B T}  \quad . 
\label{temp}
\end{equation}
The usual definition of entropy $S_{i}=k\ln(\omega_{i})$ implies that 
the extremal configuration is found where $S=S_{1}+S_{2}$ 
is extremal with its associated temperature as defined above.  
Under which circumstances can one factorize  
the dependence of $\rho$ on the total energy $E$,  
i.e. which  microcanonical distributions allow for  a 
separation of $E$ into a multiplicative 
factor? The standard way to motivate the appearance of the 
Boltzmann term in the canonical ensemble is a consequence of 
this $E$-separation  
\begin{equation}
\begin{array}{lcl}
\omega_{2}(E-E_{1})&=&\exp\bigl(\ln (\omega_{2}(E-E_{1}))\bigr)\\
&\approx& \exp\bigl(\ln(\omega_{2}(E))-
	\frac{\partial}{\partial E}\ln(\omega_{2}) \, E_{1} \bigr)\\
&\approx& \omega_{2}(E)\exp(-\beta E_{1}) \quad .
\end{array}
\label{boltzfact}
\end{equation}
The approximation in  Eq. (\ref{boltzfact}) is exact 
for $\omega_{2}(E-E_{1})$ being 
an exponential in $E$. However, this 
is not the most general way of separation.  

Now, to find the most general separation, we generalize the log function in Eq.
(\ref{boltzfact}) to some real function $f$, being twice differentiable
and with a well defined inverse $f^{-1}$.
The idea is to write  
$\omega_{}(E-E_{1})=f^{-1}\circ f \circ \omega_{}
  \bigl((E-E_*)-(E_{1}-E_*)\bigr)$
and to expand $f\circ\omega_{}$ around $E-E_*$.
If the energy $E$ is separable from the system into a factor, then  
there must exist two functions $g$ (factor) and $h$ (general Boltzmann term) 
such that 
\begin{equation}
   \omega_{}(E-E_{1})=g\bigl(\omega_{}(E-E_{*})\bigr) \,\, h(x)  \quad , 
\label{seperab}
\end{equation}
with $x:=\beta (E_{1}-E_{*})$; to simplify notation   
we write $\bar \omega:= \omega(E-E_{*})$.
Now use $f$ to find the unknown functions $g$ and $h$ by  
expanding $f\circ\omega$ to first order 
\begin{equation}
 f\bigl(\omega(E-E_1)\bigr)  
 = f\bigl(g(\bar \omega) h(x)\bigr)
 \sim   f(\bar \omega)-\bar \omega \, x \, f'(\bar \omega) \quad , 
\label{sep}
\end{equation}
which is justified for small $x$, i.e., the system being 
at or near equilibrium. 
The most general solution to this separation Ansatz is 
given by the family of equations $(f,g,h)_Q$, 
parametrized by a separation constant $Q$, 
and $C$ and $C_2$ being real constants  
\begin{equation}
 \begin{array}{lcl}
 f_{}(\omega)&=&C\, \omega^{1-Q} +C_2   \\
 g_{}(\omega)&=&\omega    \\
 h_{}(x)     &=&\bigl[1-(1-Q)x \bigr]^{\frac{1}{1-Q}}   
\end{array}
\label{result1}
\end{equation}
To prove this, set $x=0$ and $h_0=h(0)$, 
so that Eq. (\ref{sep}) yields $f(g(\bar \omega) h_0)= f(\bar \omega)$, 
which means $g(\bar \omega) =\frac{\bar \omega}{h_0}$. 
Without loss of generality set $h_0=1$,  and arrive at 
$f(\bar \omega h(x)) = f(\bar \omega) - \bar \omega x f'(\bar \omega)$. 
Form partial derivatives of this expression with respect 
to $x$ and $\bar \omega$, and eliminate the $f'(\bar \omega h)$ term  
from the two resulting equations 
\begin{equation}
\begin{array}{lcl}
f'(\bar \omega h) h'   &=& - f'(\bar \omega)  \\
f'(\bar \omega h) h    &=&  (1-x)f' -\bar \omega x f'' 
\end{array} 
\end{equation}
to  arrive at the separation equation 
\begin{equation}
1-\frac{1}{x}\left( \frac{h}{h'} +1\right)  = 
- \bar \omega \frac{f''(\bar \omega)}{f'(\bar \omega)} =Q  \quad , 
\end{equation}
where $Q$ is a mathematically necessary separation constant.  The differential 
equation $1-\frac{1}{x}\bigl( \frac{h}{h'} +1\bigr)  =Q $
is straight forwardly solved to give the general Boltzmann term, 
$h(x)=\bigl[1-(1-Q)x\bigr] ^{\frac{1}{1-Q}}$, using $h(0)=1$ 
to fix the integration constant. 
Equation $- \bar \omega \frac{f''(\bar \omega)}{f'(\bar \omega)} =Q$
means, 
$f(\bar \omega) =C_1 \frac{1}{1-Q} \bar \omega^{1-Q} +C_2$, with 
$C_1$ and $C_2$ integration constants. $f$ is 
strictly monotonous except for $Q=1$, where it is constant.  
The term of interest in the canonical distribution can now be written  
as the generalized Boltzmann factor 
\begin{equation}
\omega_{2}(E-E_{1})= 
 \omega_{2}(E-E_*) \bigl[ 1-(1-Q) \beta(E_1-E_*) \bigr]^{\frac{1}{1-Q}}
\quad .
\label{qdistrib}
\end{equation}
The usual Boltzmann factor Eq. (\ref{boltzfact}) is recovered 
as the special case in the limit $Q\to 1$.
If $\omega_{2}$ is of power-form,  $\omega_{2} \propto E^{1/1-Q}$, 
as is the case for a huge class of physical systems, 
Eq. (\ref{qdistrib}) holds exactly. 

The separation constant $Q$ is not specified at this level. 
What fixes $Q$? The choice of the physical system (Hamiltonian and 
characteristics of phase space) does. 
As an example, in \cite{us} we have shown that for a 
$N$-particle Hamiltonian with pair-potentials, 
in $D$ space dimensions   
\begin{equation}
H(x,p)= \sum_{i}^{N} \frac{p_i^2}{2m} + \sum_{i<j}^{N} |x_i-x_j|^{\alpha} 
\quad , 
%H(x,p) = \frac{p^2}{2m} + |x|^{\alpha} \quad . 
\end{equation}
the following relation, fixing the separation constant, holds exactly, 
\begin{equation}
\frac{1}{1-Q}=\frac{(\alpha+2)n}{2 \alpha}-1 \quad  . 
\label{relation}
\end{equation}
This equation establishes the connection between the interaction  
term in the Hamiltonian $\alpha$, the dimensionality of phase 
space, $n=DN$, and the separation constant $Q$. 
From Eq. (\ref{relation}) it is immediately clear that 
for large systems the separation constant is always 
$Q\to 1$, i.e. the classical Boltzmann term (\ref{boltzfact}) 
is recovered. For small systems, either due to low particle numbers 
or due to an effectively collapsed phase space dimensionality,  
a non-trivial $Q\neq 1$ is expected.

\section{On the definition of temperature in low-dimensional systems}

For large systems, the notion of equilibrium is well 
defined. However,  for systems where we would expect $Q\neq1$
the definition of temperature is not necessarily unique. 
In Eq. (\ref{rho}) the definition of $\beta$ or the 
inverse temperature was based on the extremal value of $\rho$,  
but this is not what is going on in a measurement. 
A measurement of temperature yields an {\it expected} value 
(due to averages of kinetic energy taken in the measurement process) 
and not the {\it extremal} value, i.e. the most likely one.
To construct a theory which is consistent with the 
{\it measured} temperature of the sample system, and by -- at the same time -- 
keep the extremal principle, 
one can now ask to modify the definition of $\rho \to \bar \rho$, 
such that the extremal value is obtained at  the measured temperature.
Effectively this amounts to an energy shift from equilibrium energy to 
expected energy, $E_* \to \bar E$.
Such a modification could look like 
\begin{equation}
\bar \rho(E_1) \equiv \bar \omega_1(E_1) \bar \omega_2(E-E_1) \quad ,
\end{equation}
at $E_1=\bar E_1$ and 
where $\bar \omega_i(E_i) \sim E_i^{\bar \lambda_i}$. 
The idea is to identify  $\bar E_1$ with the measured energy 
$\bar E_1 = \int_0^E d \epsilon \epsilon \rho(\epsilon)$, 
and not with the equilibrium $E_*$ as before. 
The variation $\delta \bar \rho = 0$ leads to the relation
\begin{equation} 
\frac{\bar \lambda_{1}}{\bar \lambda_{2}}=\frac{\bar E_1}{E-\bar E_1} \quad, 
\label{lamrel}
\end{equation}
which establishes the relation of the $\bar \lambda$s, which are 
of course not independent. 
Now we note that 
\begin{equation}
\bar E_1=\int_{0}^{E_1} d\epsilon \epsilon \omega_{1}(\epsilon)\omega_{2}(E-\epsilon) Z(E)^{-1} \\
        =E \beta(\lambda_{1}+1,\lambda_{2})/\beta(\lambda_{1},\lambda_{2}) \quad , 
\end{equation}
with the usual beta functions, 
$\beta(a,b) \equiv \int_{0}^{1} dx x^{a}(1-x)^{b}$.
Substituting $\bar E_1$ into Eq. (\ref{lamrel})
allows to compare the $\lambda$ exponents from the original 
system, Eq. (\ref{rho}), with  those of the energy shifted system, $\bar \lambda$, 
\begin{equation}
\frac{\bar \lambda_{1}}{\bar \lambda_{2}}=
\frac{1}{ \beta(\lambda_1,\lambda_2)/\beta(\lambda_1+1,\lambda_2) -1} \quad .
\end{equation}
If we now assume that the exponent of the sample system is 
known (for example from a Hamiltonian), 
and unchanged in the shifted system, $\lambda_1=\bar \lambda_1$, 
we observe that 
the shift from $E_* \to \bar E$, has as a consequence for 
the bath density, $\omega(E_2)  \to \omega(E_2)^z $, 
where $z= \frac{\lambda_1}{\lambda_2} \frac{\beta(\lambda_1,\lambda_2+1)}{\beta(\lambda_1+1,\lambda_2)}$.
For the usual entropies this means 
\begin{equation}
S_{1}=\ln(\omega_{1})\qquad S_{2}=\ln(\omega_{2}^{z})
\end{equation} 
with the temperatures ($k_B \equiv 1$)
$1/T_{i} = \partial S_{i}/\partial E$ so that equi-temperature is assumed 
when both systems contain the expected energy, i.e. $T_{1}=T_{2}\leftrightarrow \delta S = 0$.
Here we obtain non-additivity of entropies 
as a result of harmonizing the concept of temperature 
for small systems. 

In order to see that this gives consistent results we  compute the expected temperature
of both systems $\langle T_{1} \rangle$ and $\langle T_{2} \rangle$ in 
both, the most-probable state temperature interpretation
and in the expected-state temperature interpretation. 
Let us think of the sample system (1) as the thermometer then the
above definition implies that our thermometer shows the same 
average temperature obtained by multiple measurements whether we use the maximal or 
expected energy interpretation and 
$\langle T_{1} \rangle=(E/\lambda_{1})\beta(\lambda_{1}+1,\lambda_{2})/\beta(\lambda_{1},\lambda_{2})$
since we have chosen to modify only the exponent of the bath system. 
Now the bath  shows the  expectation of temperature, 
$\langle T_{2} \rangle=(E/\lambda_{2}w)\beta(\lambda_{1},\lambda_{2}+1)/\beta(\lambda_{1},\lambda_{2})$ with $w=1$ 
for the most-probable and $w=z$ for the expected-state temperature interpretation.
Consequently,  the expected-state temperature interpretation is consistent with the BG expectation, 
\begin{equation}
\frac{\langle T_{1} \rangle}{\langle T_{2} \rangle }=\frac{\lambda_{1}z}{\lambda_{2}}
\frac{\beta(\lambda_{1}+1,\lambda_{2})}{\beta(\lambda_{1},\lambda_{2}+1)}=1\quad .
\end{equation}
Note that the expected-temperature definition involves 
knowledge about sample  and bath system. On the other hand the 
most-probable-state temperature interpretation gives  
\begin{equation}
\frac{ \langle T_{1} \rangle}{\langle T_{2} \rangle}=\frac{\lambda_{2}}{\lambda_{1}}
\frac{\beta(\lambda_{1}+1,\lambda_{2})}{\beta(\lambda_{1},\lambda_{2}+1)}
\end{equation}
which in general will not equal unity. 
The corresponding notion of equilibrium allows to define 
the most-probable entropy merely upon the knowledge of the 
particular micro-systems separately.
Since we only actually measure the sample system this 
temperature difference can not be observed.
Averaging over multiple temperature measurements does not converge 
to an equailibrium temperature. 
The way of how to average multiple temperature measurements 
in order to obtain the equilibrium temperature in terms of 
the most-probable state therefore is {\it dual} to the manipulation 
of the bath entropy, constructed to match the expected-state interpretation.

It becomes evident that the usage of expected and most probable energies can 
both be consistently used in order to define the macroscopic notion of 
equilibrium. Which of the two possibilities is realized experimentally 
is defined by the choice of the experimentator.
Intuitively we would
think that in terms of keeping records of repeated experiments the 
expected-state temperature interpretation
seems more natural since we can directly identify averages on 
multiple measurements of the temperature
with the equilibrium temperature itself. 
Whatever procedure we fix, we can not escape modifying either, the average
on the replicas, i.e. multiple measurements, or the definition of the bath entropy. 
The asymmetric definition of the entropy  does not appear explicitly 
in the canonical variational principle, based on single particle entropies
where the same mechanisms are controlled by constraints.

\section{Conclusion}

Based on very general assumptions we have derived that 
the most general Boltzmann factor in the canonical ensemble 
is a $q$-exponential. We have shown that this
result might be relevant for systems of relatively 
small phase space, which might be realized for 
several types of complex systems. 
We comment on the  temperature definition in small systems
and discuss its consequences. 
Let us stress again, that all presented  arguments are strictly 
based on Hamiltonians and on the variational  principle, we never leave the field 
of classical statistical mechanics, except for not 
taking the  thermodynamic limit.

%%%%%%%%%%%%%%%%%%%%%%%%%%%%%%%%%%%%%%%%%%%%%%%%%%%%%%%%%%%%%%%%%%%%%%%%%%%%%
\bibliographystyle{unsrt}

\end{document}